\documentclass{article}
\usepackage{spconf,amsmath,graphicx,amssymb}
\usepackage{algorithm}
\usepackage{algpseudocode}
\usepackage{color}
\usepackage{cite}
\usepackage{subfig}
\usepackage{float}
\usepackage{nccmath}

\title{Active Privacy-utility Trade-off \\ Against a Hypothesis Testing Adversary}
\name{ Ecenaz Erdemir, Pier Luigi Dragotti and Deniz G{\"u}nd{\"u}z 
}
\address{Department of Electrical and Electronic Engineering, Imperial College London, UK}
\begin{document}
%
\maketitle
\begin{abstract}
\label{sec:abstract}

\vspace{-0.2cm}
We consider a user releasing her data containing some personal information in return of a service. We model user's personal information as two correlated random variables, one of them, called the \textit{secret variable}, is to be kept private, while the other, called the \textit{useful variable}, is to be disclosed for utility. We consider active sequential data release, where at each time step the user chooses from among a finite set of release mechanisms, each revealing some information about the user's personal information, i.e., the true hypotheses, albeit with different statistics. The user manages data release in an online fashion such that maximum amount of information is revealed about the latent useful variable, while the confidence for the sensitive variable is kept below a predefined level. For the utility, we consider both the probability of correct detection of the useful variable and the mutual information (MI) between the useful variable and released data. We formulate both problems as a Markov decision process (MDP), and numerically solve them by advantage actor-critic (A2C) deep reinforcement learning (RL).
\end{abstract}

\vspace{-0.5cm}
\section{Introduction}

\vspace{-0.2cm}
Recent advances in Internet of things (IoT) devices and services have increased their usage in a wide range of areas, such as health and activity monitoring, location-based services and smart metering. However, in most of these applications, data collected by IoT devices contain sensitive information about the users. Chronic illnesses, disabilities, daily habits, presence at home, or states of home appliances are typical examples of sensitive information that can be inferred from collected data. Privacy is an important concern for the adoption of many IoT services, and there is a growing demand from consumers to keep their personal information private. Privacy has been widely studied in the literature \cite{DiffPrivacyMarkov,statInf,SkoglundEpsPriv,BizBook,ICASSP,FunnelLimits,WIFS,TIFS,DenizTobias,KLprivacy}, and a vast number of privacy measures have been introduced, including differential privacy \cite{DiffPrivacyMarkov}, mutual information (MI) \cite{statInf, SkoglundEpsPriv,BizBook,ICASSP,FunnelLimits,WIFS,TIFS}, total variation distance \cite{Borzoo}, maximal leakage \cite{SankarAlphaMaxLeak,MaxLeak}, and guessing leakage \cite{HamedBorzoo}, to count a few.
\makeatletter{\renewcommand*{\@makefnmark}{}\footnotetext{This work was partially supported by the European Research Council (ERC) through project BEACON (No. 677854).}\makeatother}

In this paper, we consider inference privacy of user data, which refers to the
protection from an adversary’s attempt to deduce sensitive information from an underlying distribution. We specifically consider an active learning scenario for privacy-utility trade-off against a hypothesis testing (HT) adversary. In this setting, we assume that a user wants to share the ``useful'' part of her data with a utility provider (UP). However, the UP, which we will call the \textit{adversary}, might also try to deduce user's ``secret" information from the shared data. We model the user's secret and useful data as correlated discrete random variables (r.v.'s). The user's goal is to prevent the secret from being accurately detected by the adversary while the useful data is revealed to the adversary for utility. Differently from the existing works \cite{statInf,SkoglundEpsPriv,Borzoo,SankarAlphaMaxLeak,HamedBorzoo,Funnel}, which typically consider a one-shot data release problem, we consider a discrete time system, and assume that the user can choose from among a finite number of data release mechanisms at each time. These might correspond to different types of sensor readings. While each measurement reveals some information about user's latent states, we assume that each sensor has different measurement characteristics, i.e., conditional probability distributions. User's objective is to choose a data release mechanism at each time, in an online fashion, to receive maximum utility while keeping the adversary's confidence for the sensitive information below a prescribed value.

Our problem is also similar to active sequential HT \cite{TaraIU,TaraMary,Urbashi}, where the objective is to detect the true hypothesis as quickly as possible; on the contrary, our goal is to maximize the utility of the useful data while keeping the adversary's confidence about the secret below a prescribed threshold. We consider two utility measures for the useful data: the adversary's confidence level on the useful data and the MI between the useful and released data. Note that maximizing MI does not necessarily maximize the confidence on the true useful data at the time of decision; instead, it maximizes leakage not only for the true hypothesis but for all possible hypotheses.

We recast the problem under both utility measures as a partially observable Markov decision process (POMDP) and use advantage actor-critic (A2C) deep reinforcement learning (RL) framework to evaluate and optimize the utility-privacy trade-off. Finally, we provide a numerical comparison between the proposed policies studying the adversary's confidence on the true useful hypothesis under different confidence thresholds on the secret. We also compare the policies in terms of decision making time and information leakage.

The remainder of the paper is organized as follows. We present the problem formulation and the MDP approach in Sections \ref{sec:ProbForm} and \ref{sec:MDP}, respectively. MI utility is introduced in Section \ref{sec:MIpolicy}, and the evaluation through A2C deep RL is presented in Section \ref{sec:SimRes}. We conclude our work in Section \ref{sec:Conc}.

\vspace{-0.5cm}
\section{Problem formulation}
\label{sec:ProbForm}

\vspace{-0.3cm}
We consider a user that wants to share her data with a potential adversary in return of utility. The data reveals information about two underlying latent variables; one represents the user's sensitive information, called the \textit{secret}, while the other is non-sensitive useful part, and is intentionally disclosed for utility. The \textit{adversary}, in this context, can model an honest but curious UP. It can also model a third party that may illegitimately access the released data. The user's goal is to maximize the adversary's confidence for the non-sensitive information to gain utility, while keeping his confidence in the secret below a predefined level.


Let $\mathcal{S}=\{0,1, \dots, N-1\}$ and $\mathcal{U}=\{0,1, \dots, M-1\}$ be the finite sets of the hypotheses represented by the r.v.'s $S \in \mathcal{S}$ for the secret and $U \in \mathcal{U}$ for the non-sensitive useful information, respectively. Consider a finite set $\mathcal{A}$ of different data release mechanisms available to the user, each modeled with a different statistical relation with the underlying hypotheses. For example, in the case of a user sharing activity data, e.g., Fitbit records, set $\mathcal{A}$ may correspond to different types of sensor measurements the user may share. Useful information the user wants to share may be the exercise type, while the sensitive information can be various daily habits. We assume that the data revealed at time $t$, $Z_t$, is generated by an independent realization of a conditional probability distribution that depends on the true hypotheses and the chosen data release mechanism $A_t \in \mathcal{A}$, denoted by $q(Z_t|A_t,S,U)$.

The user's goal is to disclose $U$ through the released data $Z_t$, as long as the adversary's confidence in $S$ is below a certain threshold. We assume that the observation statistics $q(Z_t|A_t,S,U)$ and the employed data release mechanism $A_t$ are known both by the user and the adversary. To maximally confuse the adversary, the user selects action $A_t$ with a probability distribution $\pi(A_t|Z^{t-1}, A^{t-1})$ conditioned on the adversary's observation history up to that time, $\{Z^{t-1}, A^{t-1}\}$. If the user has the knowledge of the true hypotheses, she can select the actions depending on both the observation history and the true hypotheses. However, our assumption is that the true hypotheses are unknown to all the parties involved.

The optimal strategy for the adversary is to employ classical \textit{sequential HT}, i.e., he observes the data samples released by the user and updates its belief on the true hypotheses accordingly. We define the belief of the adversary on hypotheses $S$ and $U$ after observing $\{Z^{t-1},A^{t-1}\}$ by

\vspace{-0.2cm}
\begin{align}
    \hspace{-0.3cm}
    \beta_t(s,u)=P(S\hspace{-0.1cm}=s,U\hspace{-0.1cm}=u|Z^{t-1}=z^{t-1},A^{t-1}=a^{t-1}),
\end{align}

\vspace{-0.2cm}
\hspace{-0.5cm}where $s$, $u$, $z_t$ and $a_t$ are the realizations of $S$, $U$, $Z_t$, and $A_t$, respectively. The adversary's belief on the secret is $\beta_t(s)=\sum_{u\in \mathcal{U}}\beta_t(s,u)$.
Let $\tau$ be the time that we believe the adversary reaches the prescribed confidence threshold on the secret. The user stops releasing data at this point.
The main objective of this paper is to obtain a policy $\boldsymbol{\pi}$, which generates the best action probabilities, such that the adversary's belief on the true $U$ at time $\tau$ is maximized.
Therefore, our goal is to solve the following problem:

\vspace{-0.3cm}
\begin{align}
    &\underset{A_0,A_1,\dots,A_{\tau}}{\text{maximize}}
    && \beta_{\tau}(U)\\ 
    & \text{subject to} 
    &&\beta_t(S)\leq L_s, \forall t \leq \tau, \label{eq:orgProb}
\end{align}

\vspace{-0.3cm}
\hspace{-0.5cm}where $L_s$ is a predetermined scalar of the user's choice. Note that the trade-off between the utility and privacy will be obtained by considering a range of $L_s$ values

\vspace{-0.4cm}
\section{POMDP Formulation}

\vspace{-0.3cm}
\label{sec:MDP}
The above privacy-utility trade-off against a HT adversary can be recast as a POMDP with partially observable static states $\{S,U\}$, actions $A_t$, and observations $Z_t$. POMDPs can be reformulated as belief-Markov decision processes (belief-MDPs) and solved using classical MDP solution methods. Hence, we define the state of the belief-MDP as the adversary's belief on hypotheses $\{S,U\}$ after observing $\{Z^{t-1},A^{t-1}\}$, i.e., $\beta_t(s,u)$. After defining the states as the belief, the user's action probabilities become conditioned on the belief distribution, i.e., $\pi(A_t=a_t|\beta_t)$, while the observation probabilities are the same as before.

The user stops sharing data when the adversary's belief on any secret $s \in \mathcal{S}$ exceeds a threshold. Therefore, the problem is an episodic MDP, which ends when a final state is reached. We define a new state space $\mathcal{X}= P(\mathcal{S,U}) \cup \{F\}$ of size $N\times M$, where $P(\mathcal{S,U})$ is the belief space, and $F$ is a recurrent final state reached when the adversary's confidence on $\mathcal{S}$ surpasses the prescribed maximum value. After a single observation $\{z_t,a_t\}$, the adversary updates its belief by Bayes' rule as follows:

\vspace{-0.25cm}
\begin{align}
    \beta_{t+1}(s,u) = \frac{q(z_t|a_t,s,u)\pi(a_t|\beta_t)\beta_t(s,u)}{\sum\limits_{\hat{s},\hat{u}}q(z_t|a_t,\hat{s},\hat{u})\pi(a_t|\beta_t)\beta_t(\hat{s},\hat{u})},
    \label{BeliefUpdate}
\end{align}

\vspace{-0.35cm}
\hspace{-0.5cm}where $\beta_{t+1}(s,u)$ can also be denoted by $\phi^{\boldsymbol{\pi}}(\beta,z,a)$ in time-independent notation. Hence, the state transitions of the belief-MDP are governed by the observation probabilities of different actions, $q(z_t|a_t,s,u)$.
If $\beta_t(s) \geq {L_s}$ holds for any secret $s \in \mathcal{S}$, we transition to the final state $F$. The overall strategy for belief update is represented by the Bayes' operator as follows:

\vspace{-0.35cm}
\begin{equation}
    \phi^{\boldsymbol{\pi}}(x,z,a) \hspace{-0.1cm} = \hspace{-0.1cm}
    \begin{cases}
      \phi^{\boldsymbol{\pi}}(\beta,z,a), \hspace{-0.3cm} & \text{if}\ x=\beta(s,u) \text{ for } \beta(s) < {L_s} \\
      F, & \text{if}\ x=\beta(s,u) \text{ for } \beta(s) \geq {L_s} \\
      F, & \text{if}\ x=F.
    \end{cases} \nonumber
  \end{equation}
  
\vspace{-0.25cm}  
We define an instantaneous reward function for the current state, which induces policy $\boldsymbol{\pi}$ when maximized:

\vspace{-0.35cm}
\begin{equation}
 \hspace{-0.2cm}   r_{\beta}(x)=
    \begin{cases}
      0,        & \text{if}\ x=\beta(s,u) \text{ for } \beta(s) <  {L_s} \\
      \max\limits_u \beta(u), & \text{if}\ x=\beta(s,u) \text{ for } \beta(s) \geq  {L_s} \\
      0,        & \text{if}\ x=F.
    \end{cases} \nonumber
  \end{equation}

\vspace{-0.25cm}
Due to the belief-based utility, we call this approach belief-reward policy.
According to her strategy, the user checks if the adversary's belief on any secret exceeds a threshold ${L_s}$, if not, she believes that the adversary updates his belief as in (\ref{BeliefUpdate}) in the next time step. If the threshold is reached, the user stops data sharing, updates the state $x=\beta(s,u)$ to the final state $x=F$ and the episode ends.

We assume that the adversary follows the optimal sequential HT strategy. Since the user has access to all the information that the adversary has, it can perfectly track his beliefs. Hence, the user decides her own policy facilitating the adversary's HT strategy, episodic behavior and belief. Accordingly, reward function $r_{\beta}(x)$ is defined such that the user receives no reward until the adversary's belief on the secret reaches the prescribed threshold, at which point she receives a reward measured by the adversary's current belief on the true useful hypothesis, and the episode ends by reaching the final state.

The corresponding Bellman equation induced by the optimal policy $\boldsymbol{\pi}$ can be written as \cite{Puterman}, 

\vspace{-0.3cm}
\begin{align}
    \mathcal{V}^{\boldsymbol{\pi}}(\beta) = \hspace{-0.3cm} \max_{\pi(a|\beta) \in P(\mathcal{A})} \hspace{-0.1cm} \Big\{ r(\beta, \pi(a|\beta))+\mathbb{E}_{z,a}\mathcal{V}^{\boldsymbol{\pi}}(\phi^{\boldsymbol{\pi}}(\beta,z,a)) \Big\},
    \nonumber
\end{align}

\vspace{-0.35cm}
\hspace{-0.55cm}where $\mathcal{V}^{\boldsymbol{\pi}}(\beta)$ is the state-value function, and $P(\mathcal{A})$ is the action probability space. The objective is to find a policy $\boldsymbol{\pi}$ that optimizes the reward function. However, finding optimal policies for continuous state and action  MDPs is PSPACE-hard \cite{PSPACEhard}. In practice, to solve them by classical finite-state MDP methods, e.g., value iteration, policy iteration and gradient-based methods, belief discretization is required \cite{Tamas}. While a finer discretization gets closer to the optimal solution, it expands the state space; hence, the problem complexity. Therefore, we will use RL as a computational tool to numerically solve the continuous state and action space MDP.

\vspace{-0.55cm}
\section{MI utility}
\label{sec:MIpolicy}

\vspace{-0.35cm}
In this section, we consider a scenario where the UP is more interested in the statistics of the public information rather than its true value. Accordingly,
we consider MI as a utility measure; that is, the user wants to maximize the MI between the useful hypothesis and the observations by the time the adversary reaches the prescribed confidence level on the secret. MI is commonly used both as a privacy and a utility measure in the literature \cite{BizBook, TIFS, Funnel}

The MI between $U$ and $(Z^T,A^T)$ over time $T$ is given by

\vspace{-0.4cm}
\begin{align}
     I(U;Z^T,A^T)& =\sum\limits_{t=1}^T I(U;Z_t,A_t|Z^{t-1},A^{t-1}).\label{eq:MIchain}
\end{align}

\vspace{-0.35cm}
The MI between the useful hypothesis and the observations at time $t$ can be written in terms of the belief, action and observation probabilities as follows:

\vspace{-0.35cm}
\begin{align}
    I(U;&Z_t,A_t|\beta)
    =-\sum\limits_{s,u,z_t,a_t}q(z_t|a_t,s,u)\pi(a_t|\beta)\beta(s,u) \nonumber \\
    &\times \log \frac{\sum\limits_{\hat{s}}q(z_t|a_t,\hat{s},u)\pi(a_t|\beta)\beta(\hat{s},u)}{\beta(u)\sum\limits_{\bar{s},\bar{u}}q(z_t|a_t,\bar{s},\bar{u})\pi(a_t|\beta)\beta(\bar{s},\bar{u})}.
\end{align}

\vspace{-0.35cm}
Accordingly, the information reward gained in the current time step after taking action $a_t$, and releasing the corresponding observation $z_t$ is defined as

\useshortskip
\begin{equation}
    r_I(x)=
    \begin{cases}
       I(u;z_t,a_t|\beta),        & \text{if}\ x=\beta(s,u) \text{ for } \beta(s) \leq {L_s} \\
      0,        & \text{if}\ x=F.
    \end{cases} \nonumber
  \end{equation}
  
  \vspace{-0.2cm}
Adversary's belief is updated by $\phi^{\boldsymbol{\pi}}(x,z,a)$ as before.
This policy maximizes the leakage not only for the true hypothesis for $u$ but all possible hypotheses for $U$. For example, a policy may disclose a lot of information even if the adversary is confused between two out of many hypotheses, as he learns that the true state is none of the other possibilities.

\vspace{-0.5cm}
\section{Numerical results}
\label{sec:SimRes}

\vspace{-0.3cm}
The MDP formulation enables us to numerically approximate the optimal policy and the optimal reward using RL. In RL, an agent discovers the best action to take in a particular state by receiving instant rewards from the environment \cite{SuttonBarto}. In our problem, we assume that the state transitions and the reward function are known for every state-action pair. Hence, we use RL as a tool to numerically solve the optimization problem.

To integrate the RL framework into our problem, we create an artificial environment which inputs the user’s current state, $x_t \in P(\mathcal{S},\mathcal{U}) \cup F$, and action probabilities, $\pi(a_t|x_t)$ at time $t$, then calculates the reward, samples an observation $z_t$, and calculates the next state $x_{t+1}$ using the Bayes' operator as in (\ref{BeliefUpdate}). The user receives the experience tuple $(x_t,\pi(a_t|x_t),r_t,z_t,x_{t+1})$ from the environment, and refines her policy accordingly.

POMDPs with continuous belief and action spaces are difficult to solve numerically by using classical MDP solution methods. Actor-critic RL algorithms combine the advantages of value-based (critic-only) and policy-based (actor-only) methods, such as low variance and continuous action producing capability. Therefore, we use A2C deep RL for the numerical evaluation of our problem.


\vspace{-0.4cm}
\subsection{A2C Deep RL}
\label{sec:A2C}

\vspace{-0.1cm}
In the A2C deep RL algorithm, the actor represents the policy structure and the critic estimates the value function \cite{SuttonBarto}. In our setting, we parameterize the value function by the parameter vector $\theta \in \Theta$ as $V_{\theta}(x)$, and the stochastic policy by $\xi \in \Xi$ as $\pi_{\boldsymbol{\xi}}$. The error between the critic's estimate and the target differing by one-step in time is called temporal difference (TD) error \cite{SurveyACRL}. The TD error for the experience tuple $(x_t,\pi(a_t|x_t),z_t,x_{t+1},r_t)$ is estimated as

\vspace{-0.25cm}
\begin{align}
    \delta_t=r_t(x_t)+\gamma V_{\theta_t}(x_{t+1})-V_{\theta_t}(x_t),
\end{align}

\vspace{-0.25cm}
\hspace{-0.5cm}where $r_t(x_t)+\gamma V_{\theta_t}(x_{t+1})$ is called the TD target, and $\gamma$ is a discount factor chosen close to $1$ to approximate the Bellman equation for our episodic MDP. Instead of using the value functions in actor and critic updates, we use the advantage function to reduce the variance from the policy gradient. The advantage is approximated by TD error. Hence, the critic is updated by gradient ascent as:

\vspace{-0.25cm}
\begin{align}
    \theta_{t+1}=\theta_t+\eta_t^c \nabla_{\theta}\ell_{c}(\theta_t),
\end{align}

\vspace{-0.25cm}
\hspace{-0.5cm}where $\ell_c(\theta_t)=\delta_t^2$ is the critic loss, and $\eta_t^c$ is the learning rate of the critic at time $t$. The actor is updated similarly as, 

\vspace{-0.25cm}
\begin{align}
    \xi_{t+1}=\xi_t + \eta_t^a \nabla_{\xi}\ell_a(\xi_t),
\end{align}

\vspace{-0.3cm}
\hspace{-0.5cm}where $\ell_a(\xi_t)=-\ln(\pi(a_t|x_t,\xi_t))\delta_t$ is the actor loss and $\eta_t^a$ is the actor's learning rate.

In implementation, we represent the actor and critic by fully connected deep neural networks (DNNs) with two hidden layers.
The critic DNN takes the current state $x$ of size $N$ as input, and outputs the corresponding state value for the current action probabilities $V_{\theta}^{\xi}(x)$. The actor takes the state as input, and outputs the parameters $\xi$ for the corresponding state, where $\{\xi^1, \dots, \xi^{|\mathcal{\mathcal{A}}|}\}$ are the densities used to generate a Dirichlet distribution representing the action probabilities.

\vspace{-0.4cm}
\subsection{Simulation results}

\vspace{-0.2cm}
We train two fully connected feed-forward DNNs, representing the actor and critic, by utilizing ADAM optimizer \cite{ADAM}. Both networks contain two hidden layers with ReLU activation \cite{RELU}, and \textit{softmax} and \textit{tanh} at the output layers of the actor and the critic, respectively.

The results are presented for $N$ = $3$, $M$ = $3$, $|\mathcal{A}|$ = $3$ and $|\mathcal{Z}|$ = $21$, and uniformly distributed $S$ and $U$. The final state is reached when the adversary's belief on any $s \in \mathcal{S}$ exceeds the threshold ${L_s}$ for $L_s\in\{0.65,0.8,0.9,0.95\}$. Observation probabilities are selected such that each action distinguishes a different pair of hypotheses well for both $S$ and $U$. For example, we created a matrix with each row representing the conditional distribution of $z$ for different $(a,s,u)$ realizations. For sensor $a$ = $0$, we used $\mathcal{N}(0,\sigma_j)$ for $(s,u)$ = $\{(0,0), (0,1), (0,2), (1,0), (1,1), (1,2)\}$, $\mathcal{N}(1,\sigma_j)$ for $(s,u)$ = $(2,0)$, $\mathcal{N}(2,\sigma_j)$ for $(s,u)$ = $(2,1)$, and $\mathcal{N}(3,\sigma_j)$ for $(s,u)$ = $(2,2)$, and we normalized through the columns representing $z$. Here, $\sigma_j$'s are chosen randomly from the interval $[0.5,1.5]$ for each $(a,s,u)$ with index $j$=$\{1,.. ,N$$\times$$M$$\times$$|A|\}$. This sensor discloses $s$=$2$ case more than the other secrets. Moreover, $a$=$1$ and $a$=$2$ reveal more information for $s$=$1$ and $s$=$0$ cases, respectively. In this model, there is no perfect sensor which reveals only the useful hypothesis while giving no information about the secret.
As a benchmark, we also consider a random policy taking the actions independently of the adversary's observations and belief. We choose two random policies with action probabilities $\pi_{R1}(a)$=$[0.3,0.6,0.1]$ and $\pi_{R2}(a)$=$[\frac{1}{3},\frac{1}{3},\frac{1}{3}]$. When the belief on the secret exceeds the threshold, episode ends as before. 

\setlength{\textfloatsep}{10pt plus 1.0pt minus 2.0pt}

\begin{figure}[ht]
\centering
\includegraphics[width=8.6cm]{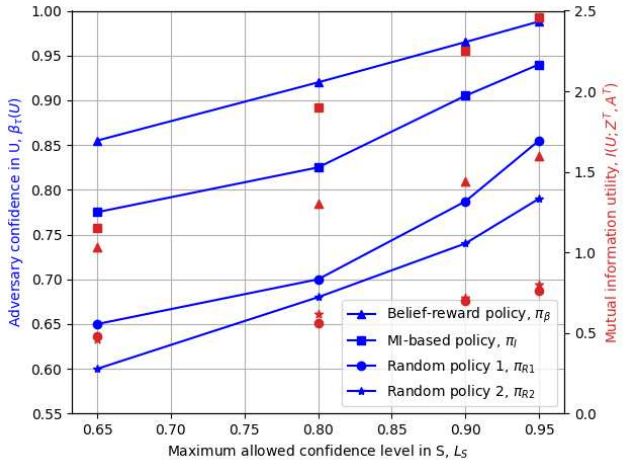}
\caption{The confidence on $U$ and MI utility w.r.t the maximum allowed confidence level on $S$ for the proposed policies.}
\label{fig:IDC_RD}
\end{figure}

In Fig. \ref{fig:IDC_RD}, we show the adversary's confidence about $U$ at the decision time on the left axis and MI between $U$ and observations on the right axis as a function of the allowed confidence level on $S$. While blue lines and red markers are scaled by the left and right axes, respectively, same markers in both colors represent the same particular policy. We represent the belief-reward and MI utility policies by, $\pi_{\beta}$ and $\pi_{I}$, respectively. We observe that through the proposed active release mechanism, the useful information can be shared with high confidence while keeping the adversary relatively confused about the secret. 
We conclude from the results that maximizing MI provides more information about the set of hypothesis $U$ than maximizing $\beta_{\tau}(U)$; however, it does not directly reveal the true hypothesis as much as $\pi_{\beta}$ reveals. However, $\pi_{I}$ still performs relatively close to the belief-reward policy $\pi_{\beta}$ for $\beta_{\tau}(U)$ at higher $L_s$.
Although the random policy provides simplicity for action selection, it has no control on the UP's confidence on the useful hypothesis. Hence, $\pi_{R1}$ and $\pi_{R2}$ perform poorly for both $\beta_{\tau}(U)$ and MI as expected since they do not use the observations to determine the best actions.

\sloppy
Note that we have not explicitly considered $\tau$ as part of our optimization. In theory, we allow unlimited time steps as long as the confidence bound on the secret is not violated. On the other hand, since the confidence level on $\mathcal{S}$ monotonically increases with time, the user stops revealing data after a finite number of steps. We observed that $\tau(L_s$= $0.65)$ = $105\pm18$, $\tau(L_s$= $0.8)$ = $250\pm46$, $\tau(L_s$= $0.9)$ = $470\pm65$ and $\tau(L_s$= $0.95)$ = $780\pm120$, which follows an increasing trend as the constraint on the secret is relaxed.
For $\pi_{I}$, we observed shorter decision times, i.e.,
$\tau(L_s$= $0.65)$ = $95\pm15$, $\tau(L_s$= $0.8)$ = $180\pm38$, $\tau(L_s$= $0.9)$ = $420\pm60$, $\tau(L_s$= $0.95)$ = $485\pm92$, which means that MI-maximizing actions also reveal more about the secret.
For $\pi_{R1}$ and $\pi_{R2}$, we observed much shorter decision times, i.e., $\tau(L_s$= $0.65)$ = $4.2\pm0.8$, $\tau(L_s$= $0.8)$ = $5.5\pm1$, $\tau(L_s$= $0.9)$ = $8.2\pm1.3$, $\tau(L_s$= $0.95)$ = $9\pm2$ and
$\tau(L_s$= $0.65)$ = $3.1\pm0.5$, $\tau(L_s$= $0.8)$ = $4.5\pm0.6$, $\tau(L_s$= $0.9)$ = $6\pm1.3$, $\tau(L_s$= $0.95)$ = $8\pm1.4$, respectively. Random policies end up choosing actions that leak significant amount of information about the secret without providing much utility.

\vspace{-0.45cm}
\section{Conclusions and Future Work}
\label{sec:Conc}

\vspace{-0.35cm}
We have seen that maximizing the MI does not necessarily reveal the true useful hypothesis with the same level of confidence as the belief-reward policy; however, this approach may be more useful when the objective is not necessarily to estimate the true value of the utility r.v., but infer its statistics. 
We have also shown that decision time is longer for higher confidence, when the good actions chosen for high utility hide the secret more than bad actions. Implementing the proposed policies on real data is in our future work plan.

\vfill\pagebreak


\bibliographystyle{IEEEbib}
\bibliography{ICASSPsub}

\begin{thebibliography}{10}

\bibitem{DiffPrivacyMarkov}
Y.~{Cao}, M.~{Yoshikawa}, Y.~{Xiao}, and L.~{Xiong},
\newblock ``Quantifying differential privacy under temporal correlations,''
\newblock in {\em 2017 IEEE 33rd Int'l Conf. Data Eng. (ICDE)}, April 2017, pp.
  821--832.

\bibitem{statInf}
F.~{du Pin Calmon} and N.~{Fawaz},
\newblock ``Privacy against statistical inference,''
\newblock in {\em 2012 50th Annual Allerton Conference on Communication,
  Control, and Computing (Allerton)}, 2012, pp. 1401--1408.

\bibitem{SkoglundEpsPriv}
A.~Zamani, T.~Oechtering, and M.~Skoglund,
\newblock ``A design framework for epsilon-private data disclosure,''
\newblock {\em ArXiv}, vol. abs/2009.01704, 2020.

\bibitem{BizBook}
E.~Erdemir, D.~G\"{u}nd\"{u}z, and P.~L. Dragotti,
\newblock ``Smart meter privacy,''
\newblock in {\em Privacy in Dynamical Systems}, Farhad Farokhi, Ed. Springer
  Singapore, first edition, 2020.

\bibitem{ICASSP}
E.~Erdemir, P.~L. Dragotti, and D.~G{\"{u}}nd{\"{u}}z,
\newblock ``Privacy-cost trade-off in a smart meter system with a renewable
  energy source and a rechargeable battery,''
\newblock in {\em IEEE Int'l Conf. on Acoustics, Speech, and Signal Processing
  (ICASSP)}, Brighton, UK, May 2019, pp. 2687--2691.

\bibitem{FunnelLimits}
B.~{Rassouli} and D.~{Gündüz},
\newblock ``On perfect privacy,''
\newblock {\em IEEE Journal on Selected Areas in Information Theory}, pp. 1--1,
  2021.

\bibitem{WIFS}
E.~Erdemir, P.~L. Dragotti, and D.~G{\"{u}}nd{\"{u}}z,
\newblock ``Privacy-aware location sharing with deep reinforcement learning,''
\newblock in {\em IEEE Workshop on Information Forensics and Security (WIFS)},
  Delft, The Netherlands, Dec 2019.

\bibitem{TIFS}
E.~{Erdemir}, P.~L. {Dragotti}, and D.~{Gündüz},
\newblock ``Privacy-aware time-series data sharing with deep reinforcement
  learning,''
\newblock {\em IEEE Transactions on Information Forensics and Security}, vol.
  16, pp. 389--401, 2021.

\bibitem{DenizTobias}
Z.~{Li}, T.~J. {Oechtering}, and D.~{Gündüz},
\newblock ``Privacy against a hypothesis testing adversary,''
\newblock {\em IEEE Trans. Inf. Forensics Security}, vol. 14, no. 6, pp.
  1567--1581, June 2019.

\bibitem{KLprivacy}
J.~{Lei} Y.-X.~{Wang} and S.~E. {Fienberg},
\newblock ``On-average kl-privacy and its equivalence to generalization for
  max-entropy mechanisms,''
\newblock in {\em Int'l Conf. Privacy in Statistical Databases}, 2016.

\bibitem{Borzoo}
B.~{Rassouli} and D.~{Gündüz},
\newblock ``Optimal utility-privacy trade-off with total variation distance as
  a privacy measure,''
\newblock {\em IEEE Trans. Inf. Forensics Security}, vol. 15, pp. 594--603,
  2020.

\bibitem{SankarAlphaMaxLeak}
J.~{Liao}, O.~{Kosut}, L.~{Sankar}, and F.~P. {Calmon},
\newblock ``A tunable measure for information leakage,''
\newblock in {\em 2018 IEEE Int'l Symp. Inf. Theory (ISIT)}, June 2018, pp.
  701--705.

\bibitem{MaxLeak}
I.~{Issa}, S.~{Kamath}, and A.~B. {Wagner},
\newblock ``An operational measure of information leakage,''
\newblock in {\em 2016 Annual Conference on Information Science and Systems
  (CISS)}, 2016, pp. 234--239.

\bibitem{HamedBorzoo}
S.~A. {Osia}, B.~{Rassouli}, H.~{Haddadi}, H.~R. {Rabiee}, and D.~{Gündüz},
\newblock ``Privacy against brute-force inference attacks,''
\newblock in {\em 2019 IEEE Int'l Symp. Inf. Theory (ISIT)}, July 2019, pp.
  637--641.

\bibitem{Funnel}
A.~{Makhdoumi}, S.~{Salamatian}, N.~{Fawaz}, and M.~{Médard},
\newblock ``From the information bottleneck to the privacy funnel,''
\newblock in {\em 2014 IEEE Information Theory Workshop (ITW 2014)}, 2014, pp.
  501--505.

\bibitem{TaraIU}
M.~{Naghshvar} and T.~{Javidi},
\newblock ``Information utility in active sequential hypothesis testing,''
\newblock in {\em Allerton Conf. on Commun., Contr., and Comput. (Allerton)},
  Sep. 2010, pp. 123--129.

\bibitem{TaraMary}
M.~{Naghshvar} and T.~{Javidi},
\newblock ``Active m-ary sequential hypothesis testing,''
\newblock in {\em IEEE Int'l Symp. Inf. Theory}, June 2010, pp. 1623--1627.

\bibitem{Urbashi}
D.~Kartik, E.~Sabir, U.~Mitra, and P.~Natarajan,
\newblock ``Policy design for active sequential hypothesis testing using deep
  learning,''
\newblock {\em 2018 56th Annual Allerton Conference on Communication, Control,
  and Computing (Allerton)}, pp. 741--748, 2018.

\bibitem{Puterman}
M.~L. Puterman,
\newblock {\em Markov Decision Processes: Discrete Stochastic Dynamic
  Programming},
\newblock John Wiley \& Sons, Inc., USA, 1st edition, 1994.

\bibitem{PSPACEhard}
C.~H. Papadimitriou and J.~N. Tsitsiklis,
\newblock ``The complexity of markov decision processes,''
\newblock {\em Mathematics of Operations Research}, vol. 12, no. 3, pp.
  441--450, 1987.

\bibitem{Tamas}
N.~Saldi, T.~Linder, and S.~Y{\"u}ksel,
\newblock {\em Approximations for Partially Observed Markov Decision
  Processes}, pp. 99--123,
\newblock Springer Int'l Publishing, Cham, 2018.

\bibitem{SuttonBarto}
R.~S. Sutton and A.~G. Barto,
\newblock {\em Reinforcement Learning: An Introduction},
\newblock The MIT Press, second edition, 2018.

\bibitem{SurveyACRL}
I.~{Grondman}, L.~{Busoniu}, G.~A.~D. {Lopes}, and R.~{Babuska},
\newblock ``A survey of actor-critic reinforcement learning: Standard and
  natural policy gradients,''
\newblock {\em IEEE Trans. Syst., Man, Cybern., Part C (Applications and
  Reviews)}, vol. 42, no. 6, pp. 1291--1307, Nov 2012.

\bibitem{ADAM}
D.~P. Kingma and J.~Ba,
\newblock ``Adam: A method for stochastic optimization,''
\newblock {\em CoRR}, vol. abs/1412.6980, 2015.

\bibitem{RELU}
V.~Nair and G.~Hinton,
\newblock ``Rectified linear units improve restricted boltzmann machines vinod
  nair,''
\newblock 06 2010, vol.~27, pp. 807--814.

\end{thebibliography}

\end{document}